\documentclass[12pt]{iopart}

\usepackage{amssymb}
\usepackage{graphicx,tabularx}
\usepackage{bm}
\usepackage{dcolumn}
\usepackage{color}
\usepackage{ulem}
\usepackage{mathptmx}
\usepackage{moreverb}
\usepackage[symbol]{footmisc}

\definecolor{darkpastelgreen}{rgb}{0.01, 0.75, 0.24}


\begin{document}
	
\title{Quantum Oscillations Evidence for Topological Bands in Kagome Metal ScV$_6$Sn$_6$}
	
\author{Guoxin Zheng$^1$, Yuan Zhu$^{1}$, Shirin Mozaffari$^2$, Ning Mao$^3$, Kuan-Wen Chen$^1$, Kaila Jenkins$^{1}$, Dechen Zhang$^{1}$, Aaron Chan$^{1}$,  Hasitha W. Suriya Arachchige$^3$, Richa P. Madhogaria$^2$, Matthew Cothrine$^2$, William R. Meier$^2$, Yang Zhang$^{3,4}$, David Mandrus$^{2,3,5}$ and Lu Li$^{1}$}
\author{}
	
\address{	
	$^1$Department of Physics, University of Michigan, Ann Arbor, MI 48109, USA\\
	$^2$Materials Science and Engineering Department, University of Tennessee Knoxville, Knoxville, Tennessee 37996, USA\\
	$^3$Max Planck Institute for Chemical Physics of Solids, Dresden 01187, Germany\\
	$^4$Department of Physics and Astronomy, University of Tennessee Knoxville, Knoxville, Tennessee 37996, USA\\
	$^5$Min H. Kao Department of Electrical Engineering and Computer Science, University of Tennessee, Knoxville, Tennessee 37996, USA\\
	$^6$Materials Science and Technology Division, Oak Ridge National Laboratory, Oak Ridge, Tennessee 37831, USA}
\ead{luli@umich.edu}
\vspace{10pt}

\begin{abstract}
Metals with kagome lattice provide bulk materials to host both the flat-band and Dirac electronic dispersions. A new family of kagome metals is recently discovered in $A$V$_6$Sn$_6$. The Dirac electronic structures of this material needs more experimental evidence to confirm. In the manuscript, we investigate this problem by resolving the quantum oscillations in both electrical transport and magnetization in ScV$_6$Sn$_6$. The revealed orbits are consistent with the electronic band structure models. Furthermore, the Berry phase of a dominating orbit is revealed to be around $\pi$, providing direct evidence for the topological band structure, which is consistent with calculations. Our results demonstrate a rich physics and shed light on the correlated topological ground state of this kagome metal. 
\end{abstract}

\section{Introduction}
The kagome lattice is an ideal platform to host topological electronic states within the strong electron correlation regime due to the special lattice geometry. The characteristics of the kagome lattice band structure include a Dirac node at $K$ point, a van Hove singularity at $M$, and flat bands over the Brillouin zone (BZ).  Depending on the electron filling degrees and interactions, a wide variety of electronic states are possible, including charge density waves (CDWs) \cite{Arachchige2022, Ortiz2019, Li2021}, spin liquid states \cite{Yan2011}, charge fractionalization \cite{Brien2010, Ruegg2011}, superconductivity \cite{Ortiz2020, Ortiz2021, Wang2013}, and newly reported topological charge order \cite{Jiang2021, ZWang2021}. Recently, the coexistence of CDW and superconductivity has been discovered in the kagome metal family $A$V$_3$Sb$_5$ ($A$=K, Rb, and Cs), which have a nonzero $\mathbb{Z}_2$ topological invariant \cite{Ortiz2020, Ortiz2021, Yin2021, Fu2021, Zhou2021,Shrestha2022}. The Dirac nodal lines and nodal loops have been identified in the CsV$_3$Sb$_5$ \cite{Hao2022}. Moreover, the detailed study of quantum oscillation spectrum in CsV$_3$Sb$_5$ signified the modification of Fermi surface (FS) topology due to the CDW order \cite{Fu2021, Ortiz2021PRX, Chapai2022, Tan2023_osc}. 

Another recently discovered kagome metal family is $R$V$_6$Sn$_6$ ($R=$ Sc, Y, Gd, Tb, Dy, Ho, Er, Tm, and Lu) compounds, which host the non-magnetic vanadium kagome lattice that also exists in CsV$_3$Sb$_5$, playing an essential role on those exotic behaviors \cite{Kang2022, Ortiz2021PRX}. Among these compounds, topological Dirac surface states have been identified in GdV$_6$Sn$_6$ and HoV$_6$Sn$_6$ \cite{Peng2021, Hu2022SA}, YV$_6$Sn$_6$ is claimed to be a topological metal \cite{Pokharel2021}. The electronic and magnetic properties of $R$V$_6$Sn$_6$ ($R=$ Tb, Dy, Ho, Er, and Tm) have been studied in Ref. \cite{Zhang2022}. Here we focus on the kagome metal ScV$_6$Sn$_6$, which is the only one showing a CDW transition in $R$V$_6$Sn$_6$ family so far \cite{Arachchige2022, Meier2023}, and its non-trivial topology has been studied by angle-resolved photoemission spectroscopy (ARPES) \cite{Sante2023, Lee2023} and x-ray scattering \cite{Korshunov2023} measurements. Despite the similarities between $A$V$_3$Sb$_5$ compounds and ScV$_6$Sn$_6$, their CDWs have different wave vectors \cite{Ortiz2020, Liang2021, Kang2022, Arachchige2022}, and ScV$_6$Sn$_6$ does not host superconductivity in the ground state \cite{Arachchige2022}. These differences and potential Dirac bands inspire us to investigate the FS topology in the ground state of ScV$_6$Sn$_6$.

In this work, we present the electrical transport and magnetic properties of ScV$_6$Sn$_6$ single crystals. The resolved quantum oscillations from the Shubnikov–de Haas (SdH) effect and de Haas–van Alphen (dHvA) effect were analyzed and compared with the modeling based on WIEN2k density functional theory (DFT) calculations, indicating a slight modification of FSs affected by the CDW order formed below 92 K. Furthermore, the analysis of quantum oscillations shows one small orbit carries non-trivial Berry phase, which is consistent with the Dirac bands resolved by the theoretical calculation. These results provide direct evidence for topologically non-trivial electrical structure in ScV$_6$Sn$_6$.

\section{Methods}
Single crystals of ScV$_6$Sn$_6$ were synthesized via a self-flux growth method  \cite{Arachchige2022}. The electrical transport measurements were carried out in the Quantum Design physical property measurement system (PPMS Dynacool-14T) and the SCM2 system with an 18 T superconducting magnet with variable temperature insert (VTI) in NHMFL, Tallahassee. The torque magnetometry measurements were also performed using capacitive cantilevers in the SCM2 system. The magnetization measurements were conducted in a Quantum Design magnetic property measurement system 3 (MPMS 3) using the Vibrating Sample Magnetometer (VSM) option. DFT calculations were performed with the WIEN2k package \cite{Schwarz2022}. The angular dependence of FS cross-sectional areas was computed via SKEAF \cite{Rourke2012}.

\section{Results and Discussion}
\begin{figure}[h]
	
	\begin{center}
		\includegraphics[width=0.8 \columnwidth ]{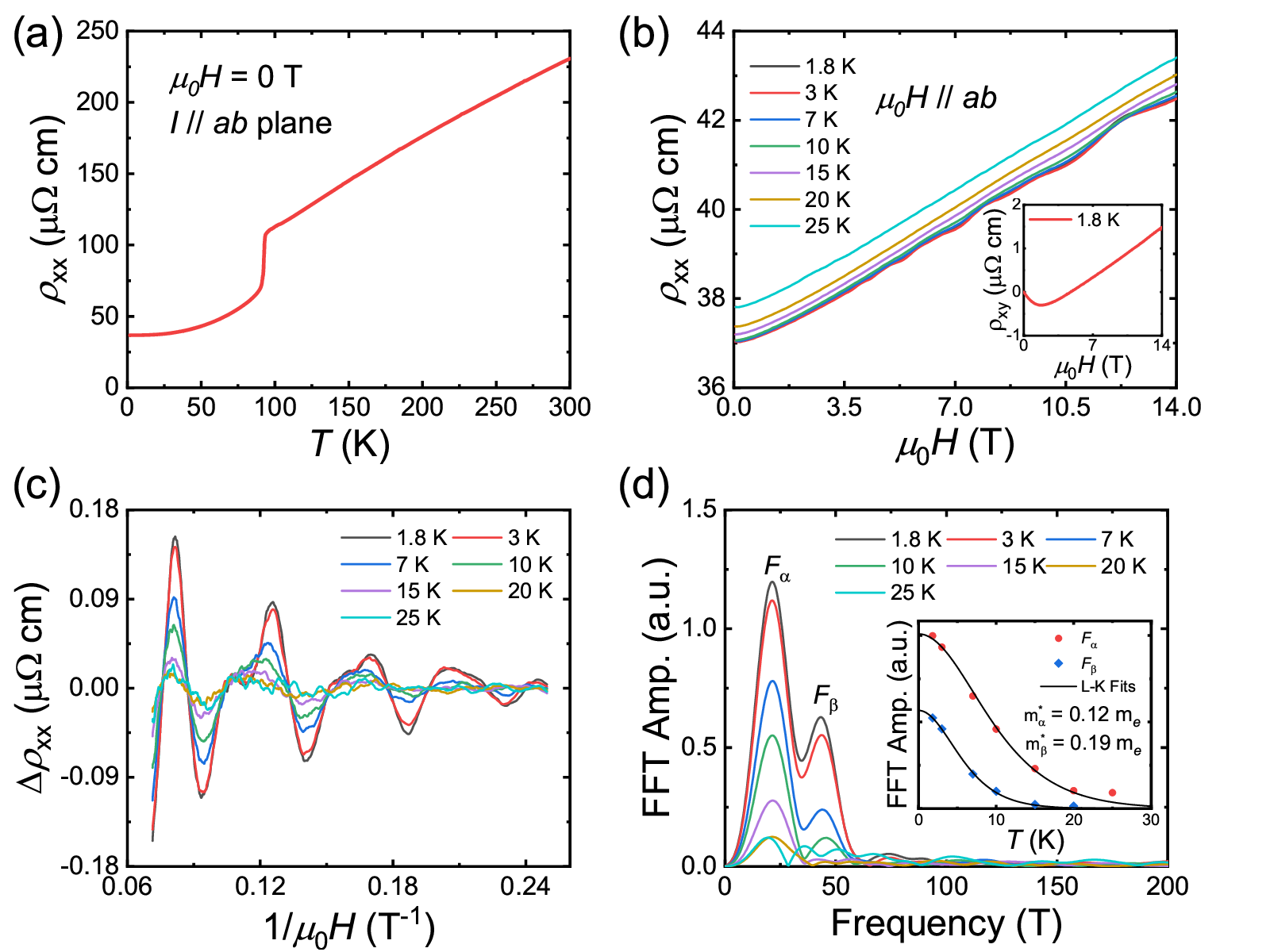}
		\caption{Electrical transport properties of ScV$_6$Sn$_6$. The current $I$ is applied in the crystalline $ ab$ plane. (a) Temperature $T$ dependence of longitudinal resistivity $\rho_{xx}$. (b) The magnetoresistance of $\rho_{xx}$ when the field $H$ is in the $ab$ plane under different $T$. The inset shows the Hall resistivity $\rho_{xy}$ measured at $T=$1.8 K, with $H$ along the $c$-axis. (c) The subtracted oscillatory patterns from (b) as a function of $1/(\mu_0H)$ under different temperatures. (d) FFT amplitude of the SdH oscillations. The inset shows the temperature dependence of the FFT amplitude  of $F_\alpha$ (red dots) and $F_\beta$ (blue dots) with LK fits to find the effective masses. } 
		\label{fig1}
	\end{center}
\end{figure}

The temperature ($T$) dependence of \textit{ab}-plane resistivity $\rho_{xx}$ is shown in figure \ref{fig1}(a). ScV$_6$Sn$_6$ resistivity exhibits a metallic behavior with a significant drop around 92 K which is confirmed to be a CDW transition \cite{Arachchige2022}, and no bulk superconductivity is observed down to 1.8 K, which is consistent with Ref. \cite{Arachchige2022}. figure \ref{fig1}(b) displays the magnetoresistance of $\rho_{xx}$ when the magnetic field $H$ is applied in the \textit{ab}-plane under different $T$. These magnetoresistance (MR) curves exhibit sub-quadratic or nearly linear behavior. This linear MR feature also reported recently in Ref. \cite{Destefano2023, Guguchia2023} could indicate an unconventional phase, such as the formation of a pseudogap as claimed in Ref. \cite{Destefano2023}. The main feature in figure \ref{fig1}(b) is the clear SdH oscillations when $H>$  3.5 T, and the amplitude decreases with increasing $T$. The transverse resistivity $\rho_{xy}$ has also been measured at $T=$ 1.8 K when the field is along the $c$ axis as shown in figure \ref{fig1}(b) inset. $\rho_{xy}$ shows a nonlinear behavior with a negative slope below 1.7 T and a positive slope above, which can be well fitted by the two-band model~\cite{Destefano2023, Shirin} (Also see supplementary material \cite{supp} section E). After subtracting a smooth non-oscillatory background, the SdH oscillations are isolated in figure \ref{fig1}(c) as a function of 1/$\mu_0H$ and visible up to 25 K. The beating periodic oscillatory patterns indicate the contribution of two different frequency components. After doing the Fast Fourier transform (FFT), the spectra are shown in figure \ref{fig1}(d), which gives two dominant frequencies, $F_{\alpha}\sim$ 21 T and $F_{\beta}\sim$ 44 T. Another orbit $F_{\gamma}\sim$ 43 T with weak amplitude showed up after we conducted the angular dependence of FFT, as shown in figure \ref{fig3}(a) top panel, which has a similar frequency with the $\beta$ orbit but comes from a different band. No high-frequency peak is observed in the FFT spectra. 

In general, the SdH oscillations with several frequencies can be considered as the linear superposition of the Lifshitz-Kosevich (LK) formula of different frequency $F$, and each LK formula can be expressed as \cite{Shoenberg1984, Murakawa2013, Fu2021}:

\begin{equation}
	\Delta \rho \propto B^{1/2} R_T R_D R_S \textup{cos}[2\pi(\frac{F}{B} - \gamma + \delta +\varphi)]
	\label{eq1}
\end{equation}

where $R_T=\frac{\alpha m^* T/B}{\textup{sinh}(\alpha m^* T/B)}, R_D=e^{-\alpha m^*T_D/B}$, and $R_S=\textup{cos}(\frac{\pi}{2}g^*m^*)$ stand for the reduction factors due to the temperature, scattering, and spin splitting. Here $T_D$ is Dingle temperature defined as $T_D=\frac{\hbar}{2\pi k_B \tau}$, where the scattering time $\tau$ is related to the electronic mobility as $\mu_e=e\tau/m^*$. $F$ is frequency, $m^*$ is effective mass in unit of free electron mass $m_e$, $\alpha=2\pi^2k_Bm_e/e\hbar=14.69$ T/K is a constant, and $g^*$ is the effective $g$ factor. The phase factor $\phi=-\gamma +\delta +\varphi$ comes from multiple factors: (1) $\varphi = \frac{1}{2}$ when $\rho_{xx} \gg \rho_{xy}$ and $\varphi = 0$ when $\rho_{xx} \ll \rho_{xy}$. (2) $\delta =0$ for a two-dimensional (2D) Fermi surface, $\delta = {-\frac{1}{8}}$ for a three-dimensional (3D) Fermi surface when the extremal orbit is the local maximum of the orbit, and $\delta ={+\frac{1}{8}}$ for a 3D Fermi surface when the extremal orbit is the local minimum of the orbit. (3), Finally, $\gamma = \frac{1}{2} - \frac{\phi_B}{2\pi}$ with $\phi_B$ the Berry phase. 

The effective masses can be estimated by fitting the oscillation amplitudes as a function of $T$ using the LK formula, shown in the inset of figure \ref{fig1}(d). The determined masses are m$_{\alpha}^*=0.12$ m$_e$ and m$_{\beta}^*=0.19$ m$_e$, both are small light pockets.

\begin{figure}[h]
	\begin{center}
		\includegraphics[width=0.8 \columnwidth ]{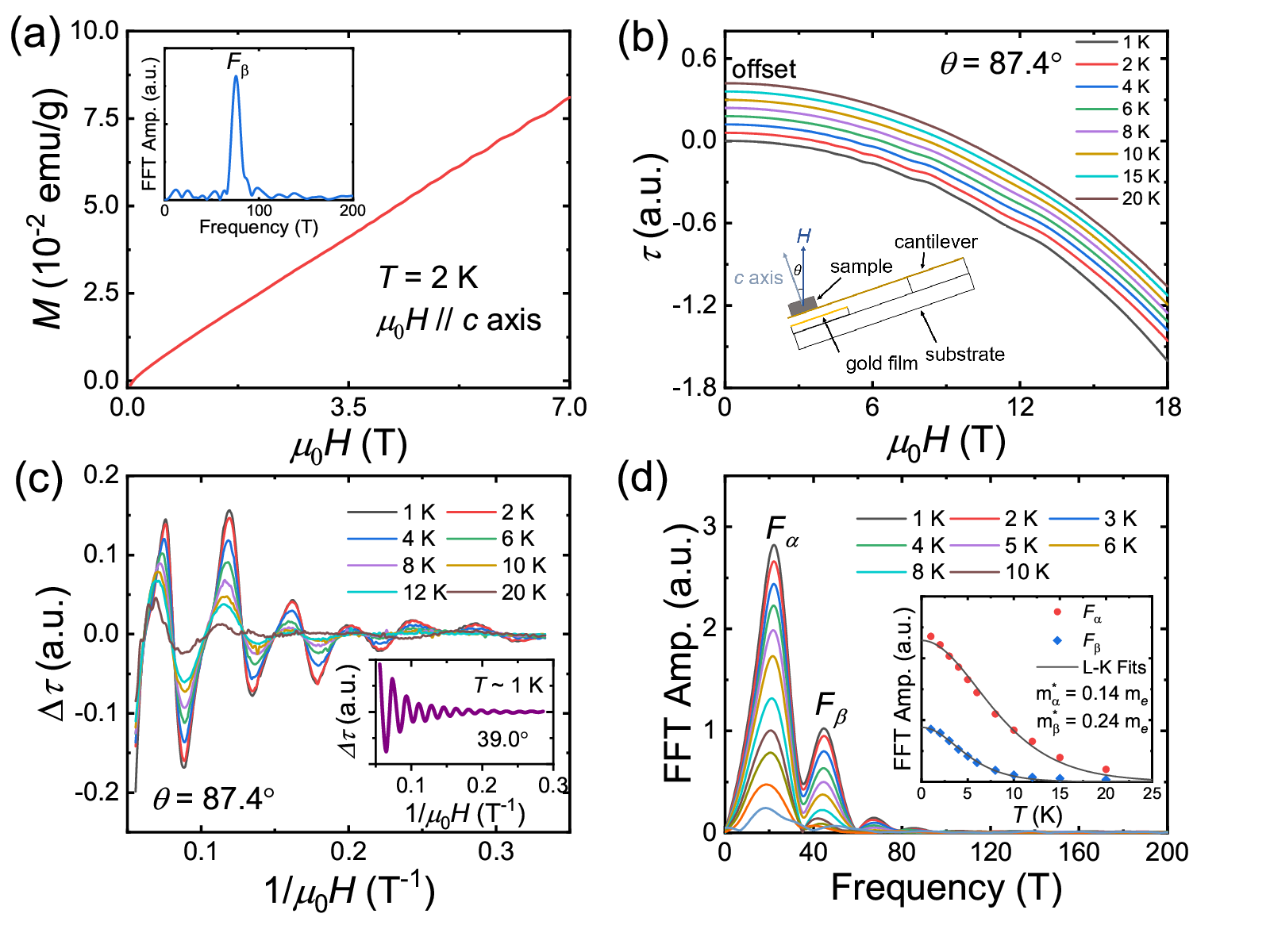}
		\caption{Magnetic property measurements in ScV$_6$Sn$_6$. (a)  The $H$-dependence of magnetization at 2 K with $H$ along the $c$ axis. The inset shows the FFT spectra of magnetization oscillations after background subtraction. (b) The $H$-dependence of torque $\tau$ at $\theta=87.4^\circ$ under different $T$. The curves have a 0.06 offset for clarity. The inset shows the cantilever torque magnetometry setup and definition of $\theta$. (c) The subtracted oscillatory patterns from (b) as a function of $1/\mu_0H$ under different $T$. Inset: quantum oscillations at 39.0$^{\circ}$ which only contain $\beta$ orbit. (d) The $T$ dependence of FFT amplitude of $F_\alpha$ (red dots) and $F_\beta$ (blue dots) at $\theta=87.4^\circ$, The effective masses are found by LK fits. } 
		\label{fig2}
	\end{center}
\end{figure}

\begin{figure}[h]
	\begin{center}
		\includegraphics[width=0.8\columnwidth ]{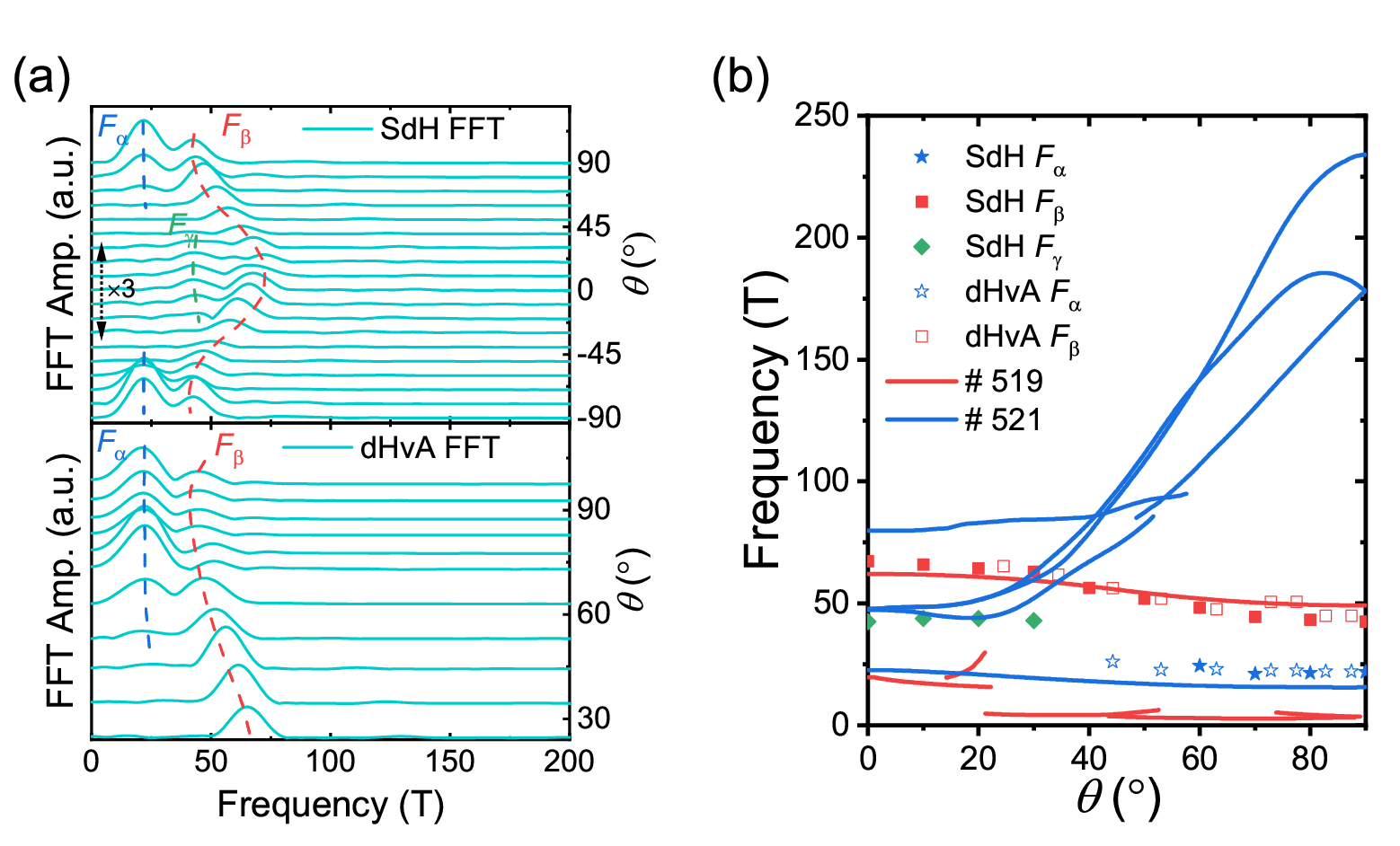}
		\caption{ Angular dependence of oscillation frequencies in ScV$_6$Sn$_6$. (a) Angular dependence of the FFT amplitudes of the SdH oscillation (up) and the dHvA oscillations (down). The spectra were shifted so that the right axis marks the tilt angle. The blue, red, and green dash curves are the guidelines to track the peak shift of $F_{\alpha}, F_\beta$ and $F_\gamma$, respectively. The FFT amplitudes in the SdH data between $|\theta|<40^\circ$ are multiplied by 3 for clarity. (b) Angular dependence of $F_{\alpha}$, $F_{\beta}$, and $F_\gamma$ are compared with our DFT calculations in the LT phase. Solid lines are calculated frequencies from bands 519 (red) and 521 (blue) from the DFT calculation shown in figure \ref{fig4}. Closed and open stars indicate $F_{\alpha}$ measured from SdH and dHvA oscillations. Closed and open squares denote $F_{\beta}$ measured from the SdH and dHvA oscillations. Closed diamonds mean $F_\gamma$ observed in the SdH data. 
		}
		\label{fig3}
	\end{center}
\end{figure}

The SdH signals are confirmed with the dHvA effect observed in the magnetization $M$ and magnetic torque of ScV$_6$Sn$_6$. The $H$-dependence of $M$ is shown in figure \ref{fig2}(a), with $H$ along $c$ axis and at $T=$ 2 K. The curve shows a typical paramagnetic response without saturation in $H$ up to 7 T. The oscillatory patterns caused by the dHvA effect appear at $H >$ 3 T, which come from a single frequency contribution resolved by the FFT spectra shown in the inset of figure \ref{fig2}(a). We identify this frequency as $F_{\beta} \sim$ 76 T with $H\parallel c$. The magnetic torque $\tau$ measurement setup is shown in figure \ref{fig2}(b) inset and $\theta$ is the tilt angle between $H$ and the crystalline $c$ axis. The $H$-dependence of torque under selected $T$ is shown in figure \ref{fig2}(b) with offset. Torque magnetometry directly measures the anisotropy of magnetic susceptibility of the sample\cite{Li2008}. All torque curves have a quadratic polynomial background, consistent with paramagnetic responses ~\cite{Li2008}. The quantum oscillations in figure \ref{fig2}(b) are separated in figure \ref{fig2}(c) after subtracting a polynomial background, showing clear periodic oscillatory patterns which are observable up to 20 K. The quantum oscillations at another angle $39.0^{\circ}$ are shown in the inset of (c), which are clearly dominated by $\beta$ orbit. 

Similar to Eq. \ref{eq1}, the LK formula describing the field and temperature-dependence of the dHvA oscillations of the magnetization along field direction is given by \cite{Shoenberg1984, Chen2018}:
\begin{equation}
	\Delta M_{||}\propto -B^{1/2}R_TR_DR_S\textup{sin}[2\pi(\frac{F}{B}-\gamma+\delta)]
	\label{eq_M}
\end{equation}
Here $R_T, R_D,R_S,\gamma$, and $\delta$ have the same definitions as in Eq. (\ref{eq1}). Take the derivative on the $\Delta M_{||}$, the of dominating part of the relative magnetic susceptibility $\chi_{||}$ provides an easier way to extract the correct phase of the dHvA oscillations, which is given by:
\begin{equation}
	\Delta\chi_{||}=\frac{d(\Delta M_{||})}{dB}\propto  B^{-3/2}R_TR_DR_S\textup{cos}[2\pi(\frac{F}{B}-\gamma+\delta)]
	\label{eq_chi}
\end{equation}

figure \ref{fig2}(d) displays the FFT spectra of dHvA oscillations under different temperatures, which is similar to the SdH FFT in figure \ref{fig1}(d). Effective masses of $F_{\alpha}$ and $F_{\beta}$ found in dHvA oscillations are given in the inset of figure \ref{fig2}(d), which are slightly higher than the results in figure \ref{fig1}(d). We also notice a small peak located at 67.5 T, which is very close to the value of $F_{\alpha}+F_{\beta}$, and the mass of this frequency is 0.32 $m_e$, also close to the sum of the mass of $\alpha$ and $\beta$ orbit. Therefore we suggest this 67.5 T peak is the magnetic breakdown of $F_{\alpha}$ and $F_{\beta}$.

Furthermore, figure \ref{fig3} shows the angular dependence of the FFT of quantum oscillations in ScV$_6$Sn$_6$. The top panel of figure \ref{fig3}(a) shows the angular dependence of the SdH frequencies derived from $\rho_{xx}$. The FFT amplitudes are multiplied by 3 when $|\theta|<40^{\circ}$ for clarity. $F_{\alpha}$ has the largest amplitude when field is in the $ab$ plane, and decays quickly when $\theta$ decreases and vanishes around $\theta=60^\circ$ eventually. However, $F_{\beta}$ can be identified in all angle ranges, reaches maxima 67.5 T when $\theta = 0^{\circ}$ and drops to minima 44 T when $\theta=\pm90^\circ$. This result implies the FS pocket related to $F_{\beta}$ is a 3D ellipsoid with the minor axis along $c$ axis. From the different behavior of $F_{\alpha}$ and $F_{\beta}$ in figure \ref{fig3}, the dominated frequency $F_{\beta}$ in the magnetization (figure \ref{fig2}(a)), and the angular dependence analysis in supplementary figure S5 \cite{supp}, we can conclude that $F_{\alpha}$ and $F_{\beta}$ are two different orbits, rather than that $F_{\beta}$ is the second harmonic frequency of $F_{\alpha}$. As we mentioned before, $F_{\gamma}$ shows up when $|\theta|<30^{\circ}$, which is confirmed by the DFT calculation shown in figure \ref{fig3}(b). Similarly, the bottom panel of figure \ref{fig3}(a) gives the angular dependence FFT of dHvA data from torque measurements, which shows similar evolution patterns of $F_{\alpha}$ and $F_{\beta}$ compared with SdH FFT results. However, $F_{\gamma}$ observed in the SdH oscillations were missing in the dHvA data. This discrepancy could be caused by the different sensitivity between transport and torque measurements. The data of angular dependence of $F_{\alpha}$, $F_{\beta}$ and $F_{\gamma}$ are summarized in figure \ref{fig3}(b).

\begin{figure}[h]
	\begin{center}
		\includegraphics[width=0.8\columnwidth ]{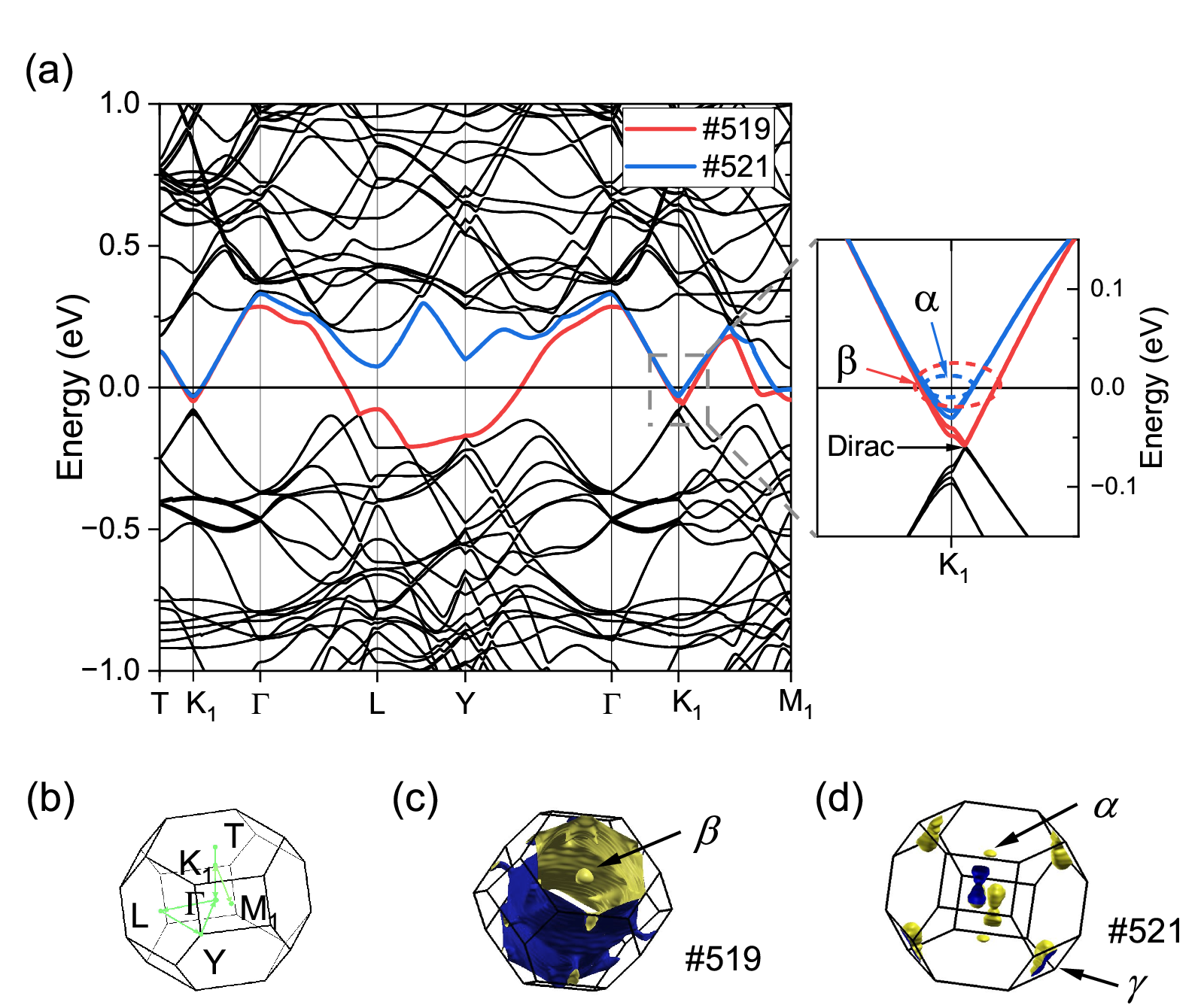}
		\caption{Electronic structure calculation in the LT phase. (a) Calculated unfolded band structure of ScV$_6$Sn$_6$ along high symmetry paths using the 50 K prime cell crystal structure. Two bands across the Fermi energy are labeled as 519 (red) and 521 (blue). The left panel is the zoom-in region around orbits and Dirac nodes. The red circle around $K_1$ indicates the $\beta$ orbit centered at $K_1$ and the blue circle means the $\alpha$ orbit is also centered at $K_1$ but from a different band. The black arrow points to the Dirac nodes along $K_1 - M_1$ in the LT phase. (b) Visualization of the BZ with labeled high symmetry points. (c) FSs of band 519. The $\beta$ orbit comes from the ellipsoid Fermi pockets located at $K_1$ point which is along the $\Gamma-T$ path.  (d) FSs of band 521. The $\alpha$ orbit can be associated with the ellipsoid shape Fermi pockets located at $K_1$ as well. $\gamma$ orbit is located at $M_1$ point with a dumbbell shape FS.}
		\label{fig4}
	\end{center}
\end{figure}

To better understand the experimental results and evaluate the influence of CDW band-folding on FSs, we performed DFT calculations based on the room temperature (RT) and low temperature (LT) crystal structures. The RT structure of ScV$_6$Sn$_6$ has the symmetry of space group $P6/mmm$, and supplementary figure S1(a) \cite{supp} shows the RT band structure considering spin-orbit coupling (SOC), which is similar to the band structure reported in Ref. \cite{Tan2023}. Multiple Dirac cones around $K$ that arise from vanadium orbitals can be seen. Next, we focus on the LT structure, which has a CDW phase transition with a (1/3, 1/3, 1/3) propagation vector ~\cite{Arachchige2022}. figure \ref{fig4}(a) shows the band structure calculation of prime cell with SOC along high-symmetry paths labeled in figure \ref{fig4}(b), and $E_F$ is shifted down 1.3 meV to match the experimental frequency results. Bands 519 (red) and 521 (blue) are the bands crossing $E_F$, and they are highlighted in figure \ref{fig4}(a), and their FSs are shown in figure \ref{fig4}(c) and (d), respectively. The left panel of (a) shows the zoom-in region around $\alpha$ and $\beta$ orbits and Dirac nodes, and we can clearly see the degenerated bands resulting from SOC. A comparison between DFT calculations and the angular dependence of the frequencies measured in SdH and dHvA oscillations is shown in figure \ref{fig3}(b). $F_\alpha$ observed in the experiments agrees with one frequency branch of the band 521 with a 6 T offset, which has an ellipsoid FS located at $K_1 (1/3, 1/3, 1/3)$ between $\Gamma$-$T$ in figure \ref{fig4}(d). The experimental data of  $F_\beta$ matched nicely with one branch of band 519, which is also an ellipsoid FS with a minor axis along $k_z$ located at $K_1$ in figure \ref{fig4}(c). The third frequency $F_{\gamma}$ can be assigned to one branch of band 521 located at $M_1 (1/2, 0 , 1/2)$ according to the angular dependence. Their are several calculated high frequency orbits in band 521 not observed in experiments. Therefore, further measurements under higher magnetic fields should be helpful to resolve these orbits in the future. In figure \ref{fig4}(a), we can identify there is a Dirac node with an ignorable gap between $K_1-M_1$ at just 57 meV below $E_F$, and surrounded by the $\beta$ pocket. This meV Dirac gap opened by SOC is small enough to generate a nontrivial Berry curvature in the $\beta$ orbit, in contrast to the trivial origin of $\pi$-phase-shifts in SOC metals, such as Bi$_2$O$_2$Se \cite{Guo2021}. After mapping the $k$ points back to RT phase, we recognize that $K_1$ and $M_1$ are the $K$ and $M$ points in the RT phase. The mapping method is discussed in supplementary section B \cite{supp}. 

The V atoms in $A$V$_3$Sb$_5$ compounds have large displacements 0.009-0.085 $\textup{\AA}$ and the reconstructed FSs are intimately related to the V orbitals \cite{Ortiz2021PRX}, especially near $K$, $M$ and $L$ points. In contrast, in ScV$_6$Sn$_6$, the vanadium atoms have much weaker displacements 0.004-0.006 $\textup{\AA}$ \cite{Arachchige2022}, which indicates the reconstruction of FSs in ScV$_6$Sn$_6$ might be weaker than $A$V$_3$Sb$_5$ compounds. ARPES measurements show that the Fermi surface significantly reconstructed at $\Gamma$ point while the V kagome bands near $K$ and $M$ remain almost unaltered after CDW transiton \cite{Lee2023}.

\begin{figure}[h]
	\begin{center}
		\includegraphics[width=0.8\columnwidth ]{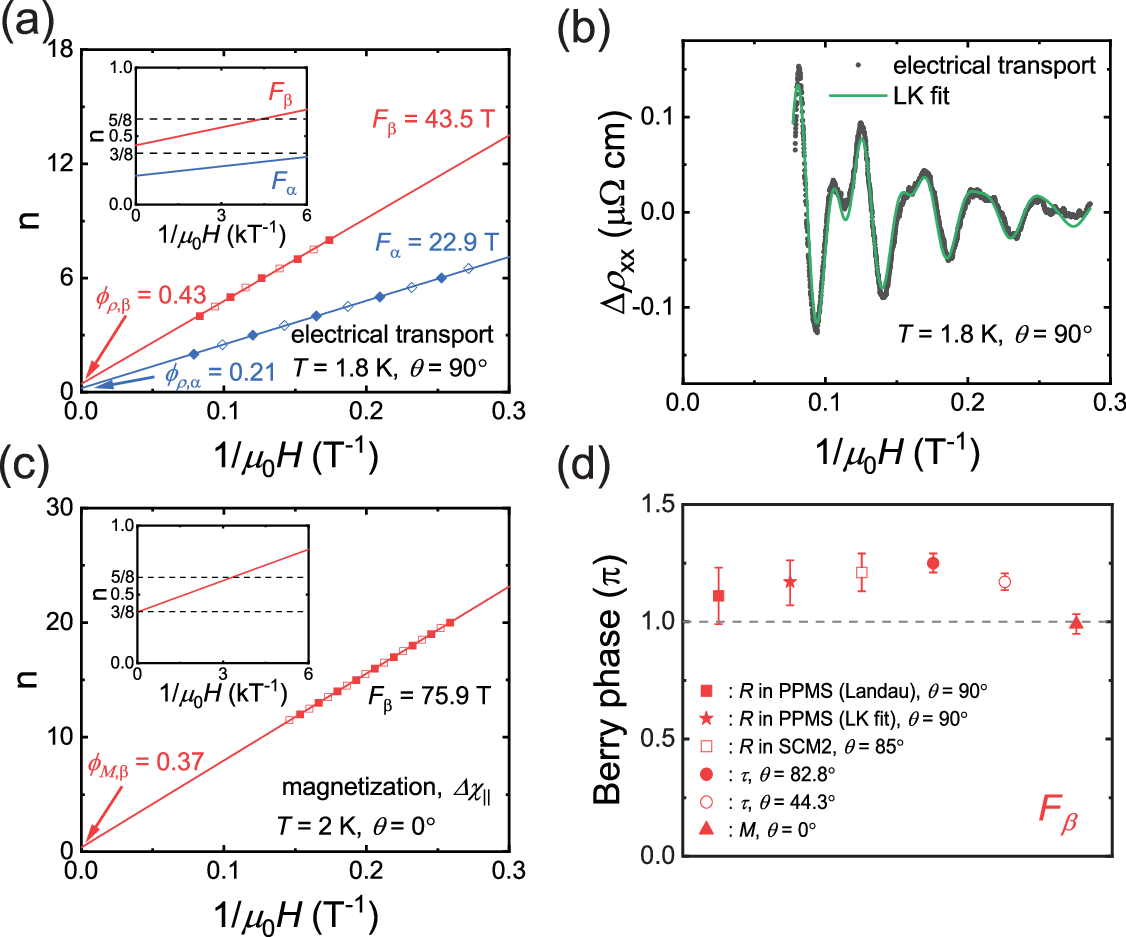}
		\caption{Berry phase identification via different methods. (a) Landau index $n$ as a function of $1/(\mu_0H)$ for $F_\alpha$ (blue squares) and $F_\beta$ (red squares) at $\theta=90^\circ$, derived from the $\rho_{xx}$ data. The linear lines are fit to the landau index. Inset is the zoom-in view around the intercept. (b) The oscillatory patterns in $\rho_{xx}$ at 1.8 K when $\theta$ is $90^\circ$. The green curve is the fit using the two-frequency LK formula. (c) Landau plot of index $n$ derived from magnetization for $F_\beta$, with the inset a zoom-in view near the intercept. (d) $\pi$ Berry phase of $\beta$ orbit resolved from both the SdH and dHvA signals in different angles. The error bars originate from the Landau index fitting or two-component LK fitting error.}
		\label{fig5}
	\end{center}
\end{figure}

Recently, Tan et al. discussed the topology of ScV$_6$Sn$_6$ \cite{Tan2023} based on the RT electronic structure. At both RT and LT phases, the Dirac cone near $K_1$ is surrounded by $\beta$ pocket, which suggests the $\beta$ orbit could be a topologically non-trivial orbit. Indeed this deduction is revealed by the Berry phase identified from the Landau level indexing of the quantum oscillation patterns. Given that $\rho_{xx}/ \rho_{xy}\sim 30$ in ScV$_6$Sn$_6$ (see figure \ref{fig1}(b)),  the maximum in $\rho_{xx}$ corresponds to the minimum of the conductivity $\sigma_{xx}$, which marks the $B$-field for each Landau level \cite{Xiong2012}. Therefore, the Landau index $n$ is assigned to the maximum of  $\rho_{xx}$ in figure \ref{fig5}(a) because the oscillation is still in the low field limit. These Landau level indexing lines determine the intercepts in the limit of $1/B = 0$. As shown in figure \ref{fig5}(a) and its inset, the intercept of $\alpha$ pocket is 0.21, $\beta$ pockets is 0.43. The topology of $\alpha$ pocket is hard to determine here, and we give a detailed discussion in supplement \cite{supp} section F. For $\beta$ pocket, this 0.43 intercept will give a Berry phase around 1.11$\pi$, which is pretty close to topological nontrivial $\pi$ Berry phase. The Berry phase of the $\gamma$ orbit is challening to determine due to its small oscillation amplitude and its beating with the $\beta$ orbit. Therefore, we will mainly focus on studying the topolgy of $\beta$ pocket here. Figure \ref{fig5}(b) shows a two-frequency LK fit for the SdH signals that also determine the Berry phases. Similar Berry phase determination was also carried out in the dHvA oscillations in both magnetizations (figure \ref{fig5}(c)) and magnetic torques (see supplementary figure S3 in \cite{supp}). The histogram in figure \ref{fig5}(d) summarizes the Berry phases for orbit $\beta$. Based on these multiple measurements, we conclude that orbit $\beta$ is topologically non-trivial. In other words, the small orbit centered at $K$ in the RT phase (along $\Gamma$-$T$ in LT phase) is topologically non-trivial, in sharp contrast to the topologically non-trivial orbits around $M$ and $H$ point in CsV$_3$Sb$_5$ \cite{Fu2021,Shrestha2022}. In supplementary figure S8 \cite{supp}, the $\pi$ Berry phase of $\beta$ orbit is also observed in YV$_6$Sn$_6$ and LuV$_6$Sn$_6$, although these two compounds do not exhibit CDW transition. The calculations show that ScV$_6$Sn$_6$, YV$_6$Sn$_6$, and LuV$_6$Sn$_6$ have nearly identical band structures at room temperature \cite{Pokharel2021, Shirin,Tan2023}. Thanks to the no change of the crystal structure, the band structures of YV$_6$Sn$_6$ and LuV$_6$Sn$_6$ should be consistent from room temperature down to low temperature. Thus, we deduce that this nontrivial $\beta$ pocket also exists in ScV$_6$Sn$_6$ at RT phase, and survives in CDW transition which is in agreement with the ARPES results \cite{Lee2023}. The discussion is further elaborated in supplement \cite{supp} section G. Our observation shows the robustness of this topological nature of the $\beta$ pocket, which indicates that this transport-detectable topological non-trivial orbit is a peculiar property shared in $R$V$_6$Sn$_6$ family, and particularly the CDW transition in ScV$_6$Sn$_6$ will not destroy this Dirac point. In addition, the intrinsic anomalous Hall effect in $\rho_{xz}$ discovered recently in ScV$_6$Sn$_6$ is another observation to suggest that a large Berry curvature arises from the non-trivial band structure \cite{Shirin}.

Usually, in kagome metals, a CDW transition generally affects the electronic structures, and thus the relationship between the CDW and the topology is rich and complicated. For example, in the well-studied kagome metal AV$_3$Sb$_5$, photoemission studies show that the CDW does not generate obvious changes in the band structure \cite{Hao2022}, yet other studies suggest that CDW opens energy gap at the Dirac cone \cite{Jiang2021}. A detailed band structure calculation points out that the CDW creates many additional band crossings \cite{Tan2023_osc}. In the kagome metal ScV$_6$Sn$_6$, we present a case that the topological orbit survives the CDW transition. A potential reason is the corresponding Fermi surface orbit at K point comes from the V atoms and does not reconstruct much after the CDW transition \cite{Lee2023, Tan2023, Kang2023}. 

\section{Conclusion}
In conclusion, quantum oscillations have been observed by electrical transport, magnetization, and torque measurements in ScV$_6$Sn$_6$. The angular dependence of oscillation frequencies is consistent with the theoretical frequencies in LT phase DFT calculations. The comparison between the LT and RT phases calculations and the comparison between the quantum oscillations among ScV$_6$Sn$_6$, YV$_6$Sn$_6$ and LuV$_6$Sn$_6$ imply the CDW transition reconstructs the FS but preserve the topological non-trivial band. The Dirac cones from the LT DFT calculations and $\pi$ Berry phase revealed for the $\beta$ orbit show a topological non-trivial electronic band structure of ScV$_6$Sn$_6$. Therefore, ScV$_6$Sn$_6$ provides a platform to study the topological electronic systems under CDW order.

It is worth mentioning that another two studies \cite{Yi2023, Shrestha2023} on quantum oscillations in ScV$_6$Sn$_6$ were reported during the submission of this manuscript. The $\beta$ orbit around 50 T is recognized as a topological nontrivial orbit in both papers, which is consistent with the key result of this manuscript.

\section*{Data availability statement}
All data that support the findings of this study are included within the article (and any supplementary files).

\section*{acknowledgments}
The work at Michigan is supported by the National Science Foundation under Award No. DMR-2004288 (transport measurements), by the Department of Energy under Award No. DE-SC0020184 (magnetometry measurements). D. M. acknowledges support from the US Department of Energy, Office of Science, Basic Energy Sciences, Materials Sciences and Engineering Division. Y.Z. acknowledge support from the start-up fund at the University of Tennessee. S. M., H.W.S. A., R. P. M.., M.C., and W. R. M. acknowledge support from the Gordon and Betty Moore Foundation’s EPiQS Initiative, Grant No. GBMF9069 to D. M.. The experiment in NHMFL is funded in part by a QuantEmX grant from ICAM and the Gordon and Betty Moore Foundation through Grant No. GBMF5305 to K.-W. C., D. Z., G. Z., A. C., Y. Z., and K. J.. A portion of this work was performed at the National High Magnetic Field Laboratory (NHMFL), which is supported by National Science Foundation Cooperative Agreement No. DMR-1644779 and the State of Florida.

\newpage

\end{document}